\documentclass[]{aastex631}
\usepackage[normalem]{ulem}
\usepackage{color} 

\received{\today}
\graphicspath{{../figures+/}}

\begin{document}

\title{RNAAS \\
Deep HST imaging favors the bulgeless edge-on galaxy explanation for the hypothetical stellar wake created by a runaway supermassive black hole  
} 

\correspondingauthor{M. Montes}
\email{mireia.montes.quiles@gmail.com}

\author[0000-0001-7847-0393]{Mireia Montes}\altaffiliation{All authors have contributed equally} \affil{Instituto de Astrof\'\i sica de Canarias, La Laguna, Tenerife, E-38200, Spain} \affil{Departamento de Astrof\'\i sica, Universidad de La Laguna}

\author[0000-0003-1123-6003]{Jorge S\'anchez Almeida}\altaffiliation{All authors have contributed equally} \affil{Instituto de Astrof\'\i sica de Canarias, La Laguna, Tenerife, E-38200, Spain} \affil{Departamento de Astrof\'\i sica, Universidad de La Laguna}

\author[0000-0001-8647-2874]{Ignacio Trujillo}\altaffiliation{All authors have contributed equally} \affil{Instituto de Astrof\'\i sica de Canarias, La Laguna, Tenerife, E-38200, Spain} \affil{Departamento de Astrof\'\i sica, Universidad de La Laguna} 



\begin{abstract}
A long linear structure recently discovered could be the stellar wake produced by the passage of a runaway supermassive black hole (SMBH) or, alternatively, a bulgeless edge-on galaxy. We report on new very deep HST imaging that seems to be in tension with the SMBH runaway scenario but is consistent with the bulgeless edge-on galaxy scenario. The new observations were aimed at detecting two key features expected in the SMBH scenario, namely, the bow shock formed where the SMBH meets the surrounding medium, and a counter stellar wake created by another binary SMBH hypothesized as part of the ejection mechanism. Neither of these two features appears to be present in the new images, as would be expected in the edge-on galaxy scenario. 

\end{abstract}



\section{Rationale and data}
\begin{figure*}[ht!] 
\includegraphics[width=0.9\linewidth]{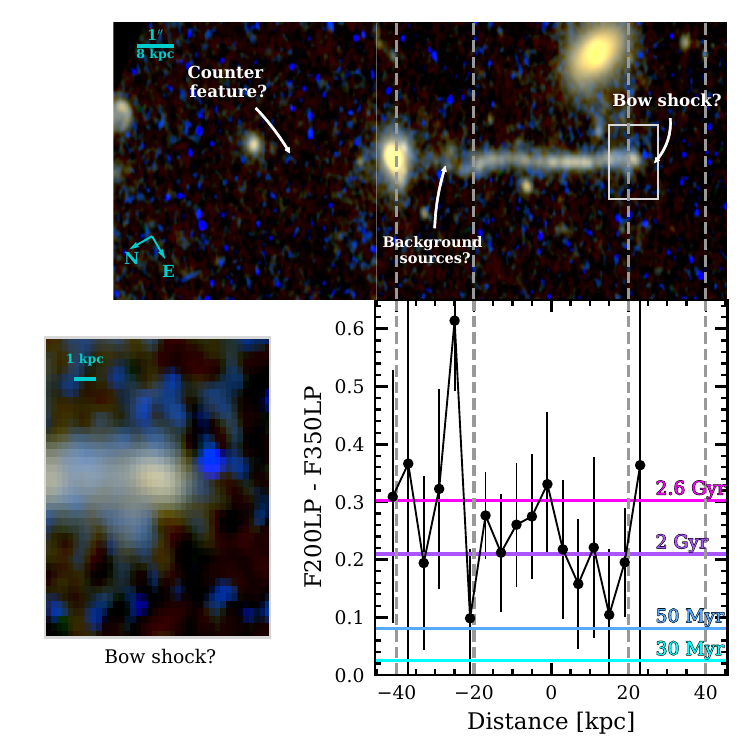}
\caption{
  Top: RGB image of the linear feature, including the field where the putative counter wake to be produced by the second pair of SMBHs is expected. 
Bottom left: Zoom-in of the box marked in the top panel, where the bow shock from the motion of the SMBH on the circumgalactic medium was expected.
Bottom right: UV emission (F200LP$-$F350LP color) along the feature averaged $\pm$1 \,kpc from its centroid. It is aligned with the image in the top panel. The plot also includes the UV F200LP$-$F350LP emission to be expected from single stellar population models (without internal dust attenuation) corresponding to various ages as labeled. The distances are referred to the dynamical center determined from the position -- velocity curve in \citet{2023A&A...673L...9S}. The vertical dashed lines, included to guide the eye, mark $\pm$20 and $\pm$40 kpc from the center of the edge-on galaxy. The blue-black knots scattered throughout the image set the background noise level.
}
\label{fig:fig1}
\end{figure*}

The  linear structure ($\sim$45\,kpc long and $\sim$1 kpc wide; Fig.~\ref{fig:fig1}, top panel) discovered by \citet{2023ApJ...946L..50V} was originally proposed to be the stellar wake triggered by the passage of a SMBH ejected from a nearby galaxy.
\citet{2023A&A...673L...9S} put forward an alternative, more conventional, explanation where the observed feature is a bulgeless edge-on galaxy observed in the restframe UV (the observed optical bands trace the UV at the redshift, $z\sim1$, of the structure).
In light of the interest sparked by the original idea, several papers have been published analyzing pros and cons of the options \citep[e.g.,][]{2023RNAAS...7...83V,2023A&A...678A.118S,2023ApJ...954L...2C,2024MNRAS.527.5503O}.

Here we report on new very deep HST imaging that appears to be in tension with the SMBH runaway scenario but is consistent with the  bulgeless edge-on galaxy scenario. The HST observation was devised to detect two key features expected in the SMBH scenario, namely, (1) the kpc-scale bow shock formed where the SMBH meets the surrounding medium to trigger star formation and (2) the counter stellar wake created by another binary SMBH expelled simultaneously with the first SMBH as part of the ejection mechanism \citep[see][]{2023ApJ...946L..50V}. The bow shock is conjectured to be a scaled-up version of those observed around runaway OB stars where they cross dense inter-stellar medium patches \citep[e.g.,][]{2017ApJ...838L..19T,2018ApJ...861...32S}. 

These two features are expected to be bolder in the far UV so the observation consists of F200LP (from $\sim$2000 to $\sim$11000\,\AA) and F350LP (from $\sim$3500 to $\sim$11000\,\AA) broad-band images with the difference (F200LP$-$F350LP) tracing the UV from 2000 to 3500\,\AA , which corresponds to the restframe far UV (from 1000 to 1800\,\AA ) at $z = 0.964$, the redshift of the structure. The individual \emph{flc} files were downloaded from the MAST archive\footnote{\url{https://mast.stsci.edu/portal/Mashup/Clients/Mast/Portal.html}} and combined using \texttt{Astrodrizzle} \citep{Fruchter2010}. The exposure times are 14\,976 s for F200LP and 14\,922 s for F350LP. The depths reached are 31.8 mag/arcsec$^2$ for F200LP and 31.3 mag/arcsec$^2$ for F350LP ($3\sigma$ in boxes of $10\arcsec\times10\arcsec)$\footnote{Compared to the 29.8 and 29.3 mag/arcsec$^2$ in the previous HST data for the F606W and F814W bands, respectively.}. All the HST data used here can be found in MAST: \dataset[10.17909/5g9g-m082]{http://dx.doi.org/10.17909/5g9g-m082}.

The top panel of Fig.~\ref{fig:fig1} shows an RGB composite of F200LP (blue), F200LP$-$F350LP (green) and F350LP (red) of the linear structure including the field where the putative counter wake to be produced by the second ejected pair of SMBHs was expected. 
The bottom left panel zooms into the box marked in the top panel, where the bow shock is expected. Finally, the bottom right panel represents the UV emission (F200LP$-$F350LP) along the feature, averaging $\pm$1\,kpc \citep[same as in][]{2023A&A...673L...9S}. The colors are corrected by the Milky Way attenuation: 0.18 and 0.10 mag for F200LP and F350LP, respectively. It also includes the UV F200LP$-$F350LP emission to be expected from the single stellar population models of \citet{Bruzual2003} using a \citet{Chabrier2003} IMF, for solar metallicity and age as labeled, without assuming any internal extinction.

\section{Results and discussion}

The new image appears to lack both key features that the HST observation aimed to detect. A blue arc-like feature identifiable with the bow shock is not obvious in the image (Fig.~\ref{fig:fig1}, bottom left panel). Similarly, a long linear feature signaling the counter wake is missing in the very deep image (Fig.~\ref{fig:fig1}, top panel). The absence of these two attributes is consistent with the edge-on galaxy explanation since they  are not expected in a regular galaxy.

The color variation (Fig.~\ref{fig:fig1}, bottom right panel) is again in tension with the stellar wake explanation. One would expect a monotonic decrease in age from left to right, i.e., from the distance closest to where the SMBH was ejected ($-$25\,kpc)  to its actual position ($+$25\,kpc). Rather, the age of the stellar wake is largest at the dynamical center of the putative edge-on galaxy as inferred from the position velocity curve \citep[][chosen to be the zero of the distance scale]{2023A&A...673L...9S} and where the youngest population should be coinciding with the SMBH ($+$25\,kpc). These ages, and the corresponding UV colors, agree with the age of the {\em host} galaxy (at $-$40\,kpc), which should be significantly older in the runaway SMBH scenario. On the other hand,  none of these properties of the color are in conflict with the edge-on galaxy alternative. Galaxies grow inside-out explaining the gradient of color, and the innermost part ($0$\,kpc,  where the bulge should be) is as old as the {\em host} galaxy.  
The reddening of the color at $-$25\,kpc deserves a special comment. In the SMBH scenario, the emission in there represents the beginning of the stellar trail and is connected to the central galaxy \citep{2023RNAAS...7...83V}. However, the actual emission emerges from knots clearly offset from the actual stellar trace (Fig.~\ref{fig:fig1}, top panel) and has a color in stark contrast with those of the linear feature. In the edge-on galaxy scenario, they could be random background galaxies reddened due to the expansion of the Universe.

A smoking-gun to distinguish between the two scenarios would involve carrying out deep imaging in the rest frame $J$ band ($\sim$1.2\,$\mu$m), roughly corresponding to the observed $K$ band. The underlaying old red stellar population should pop up if the trail is a galaxy disk whereas it should  be lacking in case of the young stellar wake resulting from the SMBH passage. This observation will settle the controversy and is easily doable with the JWST (James Web Space Telescope).

\begin{acknowledgments}
  The HST data used in this work were observed as part of the program GO-17301 (PI: van Dokkum) using the WFC3/UVIS instrument.
\noindent MM acknowledges support from the project PCI2021-122072-2B, financed by MICIN/AEI/10.13039/501100011033, and the European Union “NextGenerationEU”/RTRP and and IAC project P/302302.   
IT acknowledges support from the ACIISI, Consejer\'{i}a de Econom\'{i}a, Conocimiento y Empleo del Gobierno de Canarias and the European Regional Development Fund (ERDF) under grant with reference PROID2021010044 and from the State Research Agency (AEI-MCINN) of the Spanish Ministry of Science and Innovation under the grant PID2022-140869NB-I00 and IAC project P/302302, financed by the Ministry of Science and Innovation, through the State Budget and by the Canary Islands Department of Economy, Knowledge and Employment, through the Regional Budget of the Autonomous Community.
JSA acknowledges financial support from the Spanish Ministry of Science and Innovation, project PID2022-136598NB-C31 (ESTALLIDOS). 
\end{acknowledgments}

%

\vspace{5mm}
\facilities{HST (WFC3/UVIS)}


\software{{\tt Scipy} \citep{2020SciPy-NMeth}. 
          }



\bibliography{paper162_biblio}{}

\begin{thebibliography}{}
\expandafter\ifx\csname natexlab\endcsname\relax\def\natexlab#1{#1}\fi
\providecommand{\url}[1]{\href{#1}{#1}}
\providecommand{\dodoi}[1]{doi:~\href{http://doi.org/#1}{\nolinkurl{#1}}}
\providecommand{\doeprint}[1]{\href{http://ascl.net/#1}{\nolinkurl{http://ascl.net/#1}}}
\providecommand{\doarXiv}[1]{\href{https://arxiv.org/abs/#1}{\nolinkurl{https://arxiv.org/abs/#1}}}

\bibitem[{{Bruzual} \& {Charlot}(2003)}]{Bruzual2003}
{Bruzual}, G., \& {Charlot}, S. 2003, \mnras, 344, 1000,
  \dodoi{10.1046/j.1365-8711.2003.06897.x}

\bibitem[{{Chabrier}(2003)}]{Chabrier2003}
{Chabrier}, G. 2003, \apjl, 586, L133, \dodoi{10.1086/374879}

\bibitem[{{Chen} {et~al.}(2023){Chen}, {LaChance}, {Ni}, {Di Matteo}, {Croft},
  {Natarajan}, \& {Bird}}]{2023ApJ...954L...2C}
{Chen}, N., {LaChance}, P., {Ni}, Y., {et~al.} 2023, \apjl, 954, L2,
  \dodoi{10.3847/2041-8213/aced45}

\bibitem[{{Fruchter} \& {et al.}(2010)}]{Fruchter2010}
{Fruchter}, A.~S., \& {et al.} 2010, in 2010 Space Telescope Science Institute
  Calibration Workshop, 382--387

\bibitem[{{Ogiya} \& {Nagai}(2024)}]{2024MNRAS.527.5503O}
{Ogiya}, G., \& {Nagai}, D. 2024, \mnras, 527, 5503,
  \dodoi{10.1093/mnras/stad3469}

\bibitem[{{S{\'a}nchez Almeida}(2023)}]{2023A&A...678A.118S}
{S{\'a}nchez Almeida}, J. 2023, \aap, 678, A118,
  \dodoi{10.1051/0004-6361/202347098}

\bibitem[{{S{\'a}nchez Almeida} {et~al.}(2023){S{\'a}nchez Almeida}, {Montes},
  \& {Trujillo}}]{2023A&A...673L...9S}
{S{\'a}nchez Almeida}, J., {Montes}, M., \& {Trujillo}, I. 2023, \aap, 673, L9,
  \dodoi{10.1051/0004-6361/202346430}

\bibitem[{{S{\'a}nchez-Ayaso} {et~al.}(2018){S{\'a}nchez-Ayaso}, {del Valle},
  {Mart{\'\i}}, {Romero}, \& {Luque-Escamilla}}]{2018ApJ...861...32S}
{S{\'a}nchez-Ayaso}, E., {del Valle}, M.~V., {Mart{\'\i}}, J., {Romero}, G.~E.,
  \& {Luque-Escamilla}, P.~L. 2018, \apj, 861, 32,
  \dodoi{10.3847/1538-4357/aac7c7}

\bibitem[{{Toal{\'a}} {et~al.}(2017){Toal{\'a}}, {Oskinova}, \&
  {Ignace}}]{2017ApJ...838L..19T}
{Toal{\'a}}, J.~A., {Oskinova}, L.~M., \& {Ignace}, R. 2017, \apjl, 838, L19,
  \dodoi{10.3847/2041-8213/aa667c}

\bibitem[{{van Dokkum}(2023)}]{2023RNAAS...7...83V}
{van Dokkum}, P. 2023, Research Notes of the American Astronomical Society, 7,
  83, \dodoi{10.3847/2515-5172/acd196}

\bibitem[{{van Dokkum} {et~al.}(2023){van Dokkum}, {Pasha}, {Buzzo}, {LaMassa},
  {Shen}, {Keim}, {Abraham}, {Conroy}, {Danieli}, {Mitra}, {Nagai},
  {Natarajan}, {Romanowsky}, {Tremblay}, {Urry}, \& {van den
  Bosch}}]{2023ApJ...946L..50V}
{van Dokkum}, P., {Pasha}, I., {Buzzo}, M.~L., {et~al.} 2023, \apjl, 946, L50,
  \dodoi{10.3847/2041-8213/acba86}

\bibitem[{Virtanen {et~al.}(2020)Virtanen, Gommers, Oliphant, Haberland, Reddy,
  Cournapeau, Burovski, Peterson, Weckesser, Bright, {van der Walt}, Brett,
  Wilson, Millman, Mayorov, Nelson, Jones, Kern, Larson, Carey, Polat, Feng,
  Moore, {VanderPlas}, Laxalde, Perktold, Cimrman, Henriksen, Quintero, Harris,
  Archibald, Ribeiro, Pedregosa, {van Mulbregt}, \& {SciPy 1.0
  Contributors}}]{2020SciPy-NMeth}
Virtanen, P., Gommers, R., Oliphant, T.~E., {et~al.} 2020, Nature Methods, 17,
  261, \dodoi{10.1038/s41592-019-0686-2}

\end{thebibliography}
\bibliographystyle{aasjournal}


\end{document}